\documentclass[a4paper,11pt]{article}
\pdfoutput=1 

\usepackage{jheppub} 

\usepackage{amsthm}
\usepackage{mathrsfs}
\newcommand{\bse}{\begin{subequations}}
\newcommand{\ese}{\end{subequations}}
\newcommand{\be}{\begin{equation}}
\newcommand{\ee}{\end{equation}}
\newcommand{\bea}{\begin{eqnarray}}
\newcommand{\eea}{\end{eqnarray}}
\newcommand{\ba}{\begin{array}}
\newcommand{\ea}{\end{array}}
\newcommand{\nn}{{\nonumber}}
\newcommand{\QQ}{\mathcal{Q}}
\newcommand{\PP}{\mathcal{P}}
\newcommand{\R}{\mathcal{R}}

\newcommand{\N}{\mathcal{N}}

\newcommand{\LL}{\mathscr{L}}

\title{\boldmath Equilibrium Instability of Chiral Mesons in External Electromagnetic Field via AdS/CFT }

\author[a]{Seyed Farid Taghavi,\footnote{Corresponding author.}}
\author[a,b]{Ali Vahedi}

\affiliation[a]{School of Particles and Accelerators, Institute for Research in
	Fundamental Sciences (IPM)\\P.O. Box 19395-5531, Tehran, Iran}
\affiliation[b]{Department of physics, Kharazmi university,\\P.O.Box 31979-37551, Tehran, Iran}

\emailAdd{s.f.taghavi@ipm.ir}
\emailAdd{vahedi@ipm.ir}

\abstract{We study the equilibrium instability of chiral quarkonia in a plasma in the presence of constant magnetic and electric field and at finite axial chemical potential using AdS/CFT duality. The model in use is a supersymmetric QCD at large 't$\,$Hooft coupling and number of colors. We show that the presence of the magnetic field and the axial chemical potential even in the absence of the electric field make the system unstable. In a gapped system, a stable/unstable equilibrium state phase transition is observed and the initial transition amplitude of the equilibrium state to the non-equilibrium state is investigated. We demonstrate that at zero temperature and large magnetic field the instability grows linearly by increasing the quarkonium binding energy.	
In the constant  electric and magnetic field, the system is in a equilibrium state if the Ohm's law and the chiral magnetic effect cancel their effects. This happens in a sub-space of $(E,B,T,\mu_5)$ space with constraint equation $\sigma_B B =- \sigma E$, where $\sigma$ and $\sigma_B$ are called electric and chiral magnetic conductivity, respectively. We analyze the decay rate of a gapless system when this constraint is slightly violated.}


\begin{document} 
\maketitle
\flushbottom

\section{Introduction}

The vacuum instability in the presence of strong electric field, called \textit{Schwinger effect}, was predicted long time ago by Heisenberg and Euler and more formally developed by Schwinger \cite{Heisenberg:1935qt,Schwinger:1951nm}. As an intuitive picture, if the potential between two plates is above a critical value then the (QED) vacuum between two conducting plates sparks similar to a dielectric breakdown in a capacitor. It is well known that the decay rate of the vacuum can be obtained by $\Gamma=2\Im(\LL_{\text{eff}})$ \cite{Callan:1977pt}. Here, the  effective Lagrangian $\LL_{\text{eff}}$ is achieved by integrating out the fermionic degrees of freedom of a QED theory. The result is called Euler-Heisenberg Lagrangian and its imaginary part for a constant background electric field is \cite{Schwinger:1951nm}
\bea\label{SchwingerOrigin}
\Im(\LL_{\text{eff}})=\frac{e^2 E^2}{8 \pi^3}\sum_{n=1}^{\infty}\frac{1}{n^2}\exp\left(-\frac{n\,\pi m^2}{e E}\right),
\eea 
where $m$ is the fermion mass. The above is related to the transition amplitude of the tunneling though a potential barrier for producing a pair-particle from the vacuum. The required energy for a pair creation is $2m$ and it should be supplied by separating the particles by a distance greater than $\lambda_c$, the Compton's wavelength, via electric field. Using this picture, a critical value of the electric field can be simply estimated to be $E_{\text{cr}}\sim m^2/e$ ($\hbar=c=1$) where the exponential term in \eqref{SchwingerOrigin} becomes significantly large. In the QED theory, the 't$\,$Hooft coupling is $\lambda=g_{\text{YM}}^2\,N_c=e^2$ and therefore $E_{\text{cr}}\sim m^2/\sqrt{\lambda}$.

Recall the similarity of Schwinger effect with capacitor breakdown. The atoms of an insulator are ionized in a strong electric field and the decay rate of a band (and also Mott) insulator ground-state in a Zener breakdown is similar to the relation \eqref{SchwingerOrigin} \cite{Oka:2005}. Beside QED, we expect to observe the same phenomenon in the QCD bound states. Due to the confinement, the perturbation is not applicable in this case and non-perturbative methods such as lattice QCD or AdS/CFT correspondence are needed to study this effect in the hadrons dissociation.

In order to deal with such a non-perturbative phenomenon, we will use AdS/CFT correspondence \cite{Maldacena:1997re,Gubser:1998bc,Witten:1998qj}. According to this correspondence, the theory of $\N=4$ super Yang-Mills (SYM) is dual to the type IIB string theory on $AdS_5\times S^5$. We will use a special limit of the duality where the number of colors $N_c$ goes to infinity and 't$\,$Hooft coupling is large. In this limit, the strongly coupled $\N=4$ SYM theory is dual to the classical type IIB supergravity on $AdS_5\times S^5$. All fields in this case are in the adjoint representation of the gauge group. We add $N_f$ flavors of "quarks" in the fundamental representation by introducing $N_f$ D7-branes in the background metric. The result can be considered as a QCD-like supersymmetric theory where quarks are in a $\N=2$ hypermultiplet \cite{Aharony:1998xz,Karch:2000gx,Karch:2002sh}.

A holographic picture of the Schwinger effect in the limit $N_c\to \infty$ and large 't$\,$Hooft coupling is proposed in \cite{Semenoff:2011ng} where the pair production of "W bosons" in the $\N=4$ SYM field theory has been explored. It has been shown that demanding the reality of Dirac-Born-Infeld (DBI) action of the probe D3-brane in the $AdS_5\times S^5$ background leads to a critical value for electric field, $E_{\text{cr}}=2\pi m^2/\sqrt{\lambda}$. Note that the value of $E_{\text{cr}}$ is similar to our previous intuitive estimation up to a constant. In the context of AdS/CFT, the dissociation of the hadrons in external electric field has been studied in \cite{Hashimoto:2013mua,Hashimoto:2014dza}. In this work, the probe D3-branes in \cite{Semenoff:2011ng} are replaced by the D7-branes to add flavor quarks in the system. In fact, the DBI-action of the D7-branes plays the role of the Euler-Heisenberg Lagrangian. It is also shown that if we consider the confining force as fermion mass then the non-trivial sub-leading term in imaginary part of the D7-brane action coincides with the supersymmetric Schwinger effect. It is worth mentioning that the holographic picture of Schwinger effect in the different confining backgrounds is also studied in \cite{Ghodrati:2015rta}.

The meson spectroscopy has been studied in \cite{Kruczenski:2003be} by using the D3/D7 brane holographic model. This model is dual to the mesons in a plasma with deconfined gluons. According to this study, the meson mass can be estimated to be proportional to $m/\sqrt{\lambda}$  where $m$ stands for the quark mass. At finite temperature, above the critical value $T_c \sim m/\sqrt{\lambda}$, the  mesons are melted into the thermal gluonic plasma \cite{Hoyos:2006gb}. The gravity dual picture of the meson melting is related to the topology of the D7-branes embedding \cite{Ghoroku:2009ig,Babington:2003vm,Ghoroku:2005tf,Apreda:2005yz,Karch:2006bv,Mateos:2006nu,Albash:2006ew,Mateos:2007vn,Filev:2008xt,Evans:2008zs}. Let us mention that the meson melting might happen due to the other sources rather than temperature. Specifically, we are interested in the phase transition in the presence of axial chemical potential in the present work which has been studied in \cite{Filev:2008xt}. Using AdS/CFT correspondence, the decay width of the heavy quarkonia in a strongly-coupled thermal plasma has been studied by investigating the imaginary part of the heavy quark potential \cite{Noronha:2009da,Finazzo:2013aoa,Ali-Akbari:2014gia}.

The aim of this work is to study whether the magnetic field can dissociate mesons in a chiral supersymmetric plasma. Note that the electric field plays an active role in the meson dissociation by compensating the confining force between quarks in the meson. However, the magnetic field alone does not play such a role in the system. According to the meson spectroscopy, the mesons mass is changed by the influence of electric and magnetic field \cite{Filev:2007gb,Hoyos:2006gb,Filev:2007qu,Albash:2007bk,Albash:2007bq,Ghoroku:2009ig}. Although the meson spectrum is altered in the background magnetic field, it can not dissociate the mesons. In fact, there is a critical value for magnetic field where even at finite temperature there is no melted meson in the system in contrast with electric field \cite{Albash:2007bk,Erdmenger:2007bn,Ali-Akbari:2015bha,O'Bannon:2008bz}.

However, we can look at the effect of magnetic field on the plasma differently. Due to the chiral anomaly, there is an anomalous transport coefficient which leads to the induction of  electrical current along the magnetic field in a fluid with non-zero axial chemical potential. This effect is called chiral magnetic effect (CME),
\bea\label{CME}
\vec{J}=\frac{ \mu_5 }{2\pi^2} e\vec{B},
\eea   
and has been studied extensively in the literature \cite{Son:2009tf,Sadofyev:2010pr,Kharzeev:2004ey,Kharzeev:2007jp,Fukushima:2008xe}. Here $\mu_5$ is the axial chemical potential. One may wonder whether the CME gives rise to the instability similar to the confined mesons in the external electric field. In this article, we will see that the same instability occurs with a different mechanism. Unlike the electric field, the magnetic field is not destructive here as we expected, but, the internal energy enhancement by increasing the axial chemical potential above a critical value will liberate the confined quarks. More precisely, an equilibrium state of mesons in the background magnetic field and axial chemical potential (and temperature) above a critical value decays to a system with non-zero electric current. The critical values of axial chemical potential and temperature can be depicted in stable/unstable equilibrium phase diagram which exhibits similar features compared to the  $T$-$\,\mu_5$ phase diagram. This has already been studied by using Nambu-Jona-Lasinio model with polyakov loop (PNJL model) \cite{Fukushima:2010fe}, linear sigma model coupled to quarks and the Polyakov loop ($\text{PLSM}_q$) \cite{Chernodub:2011fr} and also lattice QCD simulation \cite{Ruggieri:2011xc}. It is worth noting that studying the full time evolution of the equilibrium state to a steady state with non-zero electric current is beyond the scope of this paper. Instead, the decay rate of the initial state will be considered here.

The AdS/CFT setup we will use is developed in \cite{Filev:2008xt,Evans:2008zs,O'Bannon:2008bz,Hoyos:2011us,Karch:2007pd}. In this setup, the $U_{\R}(1)$, the $\R$-symmetry of the $\N=2$ supersymmetry in the field theory, is assumed as the axial symmetry and the angular velocity of the rotating D7-branes around the D3-branes is dual to the axial chemical potential. In \cite{Hoyos:2011us} this picture is used to  investigate the CME where the $U_{\R}(1)$ is broken explicitly by quark mass. The effect of anisotropic plasma on CME \cite{Ali-Akbari:2014nua} and stringy corrections on the holographic picture of CME \cite{AliAkbari:2012if} are also studied in this setup. Here, we investigate the decay rate of the plasma instability by studying the imaginary part of the D7-brane action. It is interpreted as the Euler-Heisenberg Lagrangian such that the axial chemical potential is encoded in it.

Our paper is organized as follows: In section\;\ref{Review}, we review the holographic picture of supersymmetric QCD with flavor using D3/D7 brane construction. In section\;\ref{chiralInstSec}, we study the equilibrium instability for massless quarkonia and also stable/unstable equilibrium phase and its instability for a gapped system. The instability of gapless mesons in the presence of electric and magnetic field and axial chemical potential is investigated in section\;\ref{EBMUT}. Finally, in section\;\ref{summar}, we summarize the main points of the paper.

\section{A Brief Review on the Holographic picture of Supersymmetric QCD}\label{Review}

Using D3/D7 branes, a holographic description of a QCD-like field theory has been studied extensively in the past years \cite{Karch:2002sh,Grana:2001xn,Bertolini:2001qa} (for review see \cite{Erdmenger:2007cm,WiedemannBook}). In this section, we briefly review the gravity picture of the supersymmetric field theory at finite temperature and axial chemical potential where the external electric and magnetic field are present\footnote{In this case supersymmetry is broken explicitly.}. After that, we study the response of the system to the electric and magnetic field by using AdS/CFT dictionary.

Consider $N_c$ stack of D3-branes and $N_f$ stack of the D7-branes embedded in the ten dimensional space in the following form,
\bea
\begin{array}{cccccccccccc}
	& 0 & 1 & 2 & 3 & 4 & 5 & 6 & 7 & 8 & 9 \\
	D3 & \times & \times & \times & \times &  &  &  &  &  &\\
	D7 & \times & \times & \times & \times & \times  & \times  & \times  & \times  &  &\\
\end{array}.
\eea
In the limit $N_c\to\infty$ and for the large 't$\,$Hooft coupling, the D3-branes are replaced by the background metric $AdS_5\times S^5$ and a self dual R-R five form field $ F_5=dC_4$. The background metric is protected against the D7-brane back-reaction since we are in the probe limit $N_f\ll N_c$. The dual field theory of this supergravity picture is the $\N=4$ super Yang-Mills (SYM) theory in adjoint representation coupled to a $\N=2$ hypermultiplet in the fundamental representation.

In the low energy limit, the action of the D7-brane for a generic background metric can be written as
\bea\label{actionColli}
S_{D7}=S_{DBI}+S_{WZ},
\eea
where the DBI and Chern-Simons (CS) actions are given by
\begin{subequations}
	\begin{alignat}{3}
	S_{DBI} &= -N_f T_{D7} \int d^8\xi\sqrt{-\text{det}(g_{ab}^{D7}+2\pi\alpha'F_{ab})}, \\
	S_{CS} &= N_f T_{D7}\int P[\Sigma C^{(n)}]e^{2\pi\alpha'F_{ab}}.
	\end{alignat}
\end{subequations}
In the above, $g_{ab}^{D7}$ is the induced metric on the D7-brane and $T_{D7}=(2\pi)^{-7}\,g_s^{-1}\,\alpha'^{-4}$ is the tension of the brane. The symbol $P[\cdots]$ stands for the pullback of the form field $C^{(n)}$. Moreover, the field strength of the $U(1)$ gauge field living on the brane is shown by $F_{ab}$.

In the absence of the probe D7-branes, the $SO(6)$ rotational symmetry in the transverse direction of the D3-branes corresponds to $SO_{\R}(6)$, the $\R$-symmetry of the $\N=4$ SYM. The D7-branes break $SO(6)$ into a rotational symmetry in the $4567$-plane and $89$-plane which is $SO(4)\times U(1)$. This isometry corresponds to $SU_{L}(2)\times SU_{\R}(2)\times U_{\R}(1)$ where $SU_{L}(2)$ is a global symmetry and $SU_{\R}(2)\times U_{\R}(1)$ is the $\N=2$ $\R$-symmetry. 
For the separated D3- and D7-branes, the rotation in $89$-plane is also broken and correspondingly the $U_{\R}(1)$ is broken explicitly in the field theory side. 

The $\N=2$ hypermultiplet is in the fundamental representation of $SU(N_c)$. This hypermultiplet consists of two chiral superfield $Q$ and $\tilde{Q}$ with opposite chirally in the $\N=1$ notation. The two Weyl fermions of chiral superfields, $\psi$ and $\tilde{\psi}$ are transformed under $U_{\R}(1)$ with $+1$ and $-1$ charges and we can combine these two
Weyl fermions into a Dirac fermion $\Psi$. In this picture, $U_{\R}(1)$ plays the role of the axial symmetry $U_A(1)$. It is worth noting that unlike QCD there are some charged mesons under $U_A(1)$ transformation \cite{Erdmenger:2007cm}.

Here we are interested in a system at finite temperature, finite axial chemical potential and in the presence of external electric and magnetic field. For those reasons, the brane construction by the following design is chosen,
\begin{itemize}
	\item In order to have finite temperature in the dual field theory, the AdS-Schwarzschild is chosen for the background metric in the following coordinate,
	\bea \label{metric}
	ds^2&=&-|g_{tt}|dt^2+g_{xx}d\vec{x}^2 + g_{uu}du^2+g_{\theta\theta}d\theta^2+g_{\phi\phi} d\phi^2+g_{SS} ds_{S^3}^2,
	\eea 
	where
	\begin{equation} 
	\begin{split}
	g_{tt} &=\frac{L^2}{u^2}b_h(u), \quad g_{xx}=\frac{L^2}{u^2},\quad g_{uu}=\frac{L^2}{u^2}b^{-1}_h(u), \\
	g_{\theta\theta}&=L^2,\quad g_{\phi\phi}=L^2 \sin^2\theta,\quad g_{SS}=L^2 \cos^2\theta.
	\end{split}
	\end{equation}
	In the above, $b_h(u)=\left(1-u^4/u_h^4\right)$
	and $ds_{S^3}^2$ is the metric of a unit 3-sphere. Here, the boundary of the AdS is at $u\to 0$ and $u_h$ indicates the location of the horizon of the black hole. The Hawking temperature, which is also the temperature of the dual field theory, is given by
	\bea
	T=\frac{1}{\pi u_h}.
	\eea
	\item We mentioned earlier that the rotational symmetry in the $89$-plane corresponds to a global axial $U_A(1)$ symmetry in the field theory side and the length of the strings stretched between D3- and D7-branes corresponds to the mass of the quarks in the fundamental representations. 
	
	Moreover, it is shown in \cite{Das:2010yw} that the rotating D7-branes around D3-branes with constant angular velocity leads to the complex mass $m\,e^{i\phi}$ in the dual field theory where $\phi$ is the azimuthal coordinate in the $89$-plane in our case. Now it is simple to argue that the angular velocity $\partial_t\phi=\omega=2\mu_5$ of the D7-brane corresponds to the axial chemical potential $\mu_5$ in the dual field theory due to the chiral anomaly \cite{Hoyos:2011us}.
	
	\item We are interested in finding out the response of the system to the background electric and magnetic field. For simplicity, we assume $E$ and $B$ are parallel when both exist in the background. Thus, we turn on the $U(1)$ gauge field on the D7-brane in the following form,
	\bea \label{gaugeFields}
	F_{tz}=E,\quad F_{xy}=B,\quad F_{uz}=\partial_u A_z.
	\eea
\end{itemize}
The above brane setup fulfills our desires in the dual field theory.

The embedding of the D7-branes can be understood more easily in the coordinate 
\bea \label{OldCoord}
r^2+R^2=\frac{1}{u^2},\quad \frac{R}{r}=\tan\theta
\eea
where $u$ and $\theta$ are defined in the metric \eqref{metric}\footnote{The $AdS_5\times S^5$ in this coordinate is written as $$ds^2=\frac{\rho^2}{L^2}\left(-dt^2+d\vec{x}^2\right)+\frac{L^2}{\rho^2}\left(dr^2+r^2 ds_{S^3}^2+dR^2+R^2d\phi^2\right), $$ where $\rho^2=r^2+R^2$.}. The $S^5$ metric in this coordinate is $ds_{S^5}^2=dR^2+R^2d\phi^2+r^2ds_{S^3}^2$.  The $89$-plane is parametrized by polar coordinate $(R,\phi)$ where $R$ is the distance of the D3- and D7-branes. In the massless limit, we have $R(r)\to 0$ which corresponds to $\theta(u)\to0$. For more details of the D7-brane embedding in this coordinate please refer to \cite{Hoyos:2011us,WiedemannBook}. We use it in the future occasionally.

Let us go back to the coordinate mentioned in \eqref{metric}. The D7-branes are extended along $AdS_5\times S^3$ where $S^3\subset S^5$. In the static gauge, the D7-branes fill the coordinates $\{t,x_i,u,S^3\}$. Nevertheless, the embedding is not completely determined unless we specify the coordinates $\{\phi,\theta\}$ which is function of $\{t,x_i,u,S^3\}$ in general. However, translational symmetry in $x^{\mu}$ direction and rotational symmetry in the $S^3$ restrict the functionality of the remaining coordinates to the form $\theta(u)$ and $\phi(t,u)=\omega t +\varphi(u)$. We could choose  $\phi(t,u)=\omega t$ for the later ansatz. However, this choice leads to instability of the D7-brane action while the linear velocity at some point of the rotating brane may exceed the local speed of light. The additional term $\varphi(u)$ let the brane to "twist" in $89$-direction and cure this problem \cite{Hoyos:2011us}.

In this section, we focus on the massless quark limit where $\theta(u)=0$. Later, we will generalize it to the massive quarks by considering certain assumptions. A straightforward calculation leads to the following form for the DBI action,
\bea \label{SimplDBI}
S_{DBI}=-\N \int du \sqrt{\QQ_1+\QQ_2  A_z'^2},
\eea
where prime superscript means derivative with respect to $u$. The terms $\QQ_1$ and $\QQ_2$ are given by
\bea 
\QQ_1&=&\frac{L^{10}}{u^{10}}\left(1+(2\pi\alpha')^2B^2\frac{u^4}{L^4}\right)\left(1-(2\pi\alpha')^2 E^2\frac{u^4}{L^4 b_h(u)}\right),\label{Q1}\\ \QQ_2&=& \frac{L^{6}}{ u^6}b_h(u)\left(1+(2\pi\alpha')^2 B^2\frac{u^4}{L^4}\right)(2\pi\alpha')^2.\label{Q2}
\eea
In the above, we have performed the integration over $\{t,x_i,S^3\}$ trivially. Moreover, we have re-defined $S_{DBI}/V_{\mathbb{R}^{1,3}}\to S_{DBI}$ where $V_{\mathbb{R}^{1,3}}$ is the volume of the Minkowski space-time. Recalling that the 3-sphere volume is $2\pi^2$ and $L^4=\lambda \alpha'^2$, the constant behind the integral is given by
\bea
\N=\frac{\lambda N_c N_f}{L^5(2\pi)^4}.
\eea

The angular velocity $\omega$ of rotating D7-brane does not appear in the action \eqref{SimplDBI} which means that the effect of the axial chemical potential is not encoded in the DBI action and it appears in the CS action. The self-dual five form is $F_5=4 (\Omega_{S^5}+\star\Omega_{S^5})$ where $\Omega_{S^5}$ is the volume form of a 5-sphere with radius $L$. This field is found by the following R-R four form potential,
\bea\label{fourFrom}
C_4=\frac{L^4}{u^4}\;dt\wedge dx\wedge dy\wedge dz-L^4 \cos^4\theta \; d\phi \wedge \omega_{S^3},
\eea
where $\omega_{S^3}$ is the volume form of a 3-sphere with unit radius. Using above and gauge field \eqref{gaugeFields}, the CS action can be written as
\bea\label{SimplCS}
S_{CS}=-\N \int du \, \left(\PP_1\, A'_z+\PP_2\,\varphi'\right),
\eea
where
\bea \label{PPdefini}
\PP_1=L(2\pi\alpha')^2\,B\,\omega,\qquad \PP_2= L(2\pi\alpha')^2\,B\,E,
\eea
and we have performed a similar action re-definition $S_{CS}/V_{\mathbb{R}^{1,3}}\to S_{CS}$.

From \eqref{SimplDBI} and \eqref{SimplCS}, one can see that the $S_{D7}$ only depends on the derivative of $\varphi$ and $A_z$. Hence, by performing two successive Legendre  transformation, we can replace $\varphi$ and $A_z$ by two corresponding constants of motion. The constant of motion correspond to $\varphi$ is $\frac{\delta S_{D7}}{\delta \varphi'}=-\N\PP_2$ and Legendre transformation trivially cancels the term $\PP_2 \varphi'$ in the action. It means that there is no constraint on the field $\varphi$ and we can freely set it equal to zero in the massless limit. However, the constant of motion related to $A_z$ is not trivial. If we define 
\bea \label{betaDefi}
j\equiv \frac{\delta S_{D7}}{\delta A'_z}=-\frac{A'_z\QQ_2}{\sqrt{\QQ_1+\QQ_2 A'^2_z}}-\PP_1,
\eea
then the following transformation replaces field $A_z$ by its corresponding constant of motion $j$,
\begin{equation}\label{LegendreAction}
\begin{split}
\hat{S}_{D7}&=S_{D7}-\int du\, j A'_z \\
&= -\N\int du\,\sqrt{\QQ_1\left(1-\frac{\left(j/\N+\PP_1\right)^2}{\QQ_2}\right)}.
\end{split}
\end{equation}

In the next section, we will extensively explore the imaginary part of the brane action. However, we are dealing with a steady state situation here. As a result, in this case the action should be real. The reality condition of the above action demands that the quantity $\QQ_1$ and terms inside the round bracket change their sign at the same $u$. The term $\QQ_1$ changes its sign at 
\bea \label{horLikePoint}
u_{*}=\left[(2\pi \alpha' \,E)^2/L^4+\pi^4T^4\right]^{-1/4},
\eea
while it is clear from \eqref{Q2} that $\QQ_2$ is always a positive quantity. To avoid the imaginary action, the term in the round bracket inside the square root should be equal to zero at $u_{*}$. It leads to
\bea
j=-\N\PP_1-\N\sqrt{\QQ_2(u_*)}.
\eea
Using holographic renormalization, it has been shown that the electrical current in the field theory side is proportional to $j$, more precisely, $\langle J_z\rangle=-j$ \footnote{According to AdS/CFT dictionary,
$$\langle J_z\rangle=\lim_{\epsilon\to 0}\left(\frac{L}{\epsilon}\right)^4\frac{1}{\sqrt{-\gamma}}\frac{\delta S_{D7}^{\text{sub}}}{\delta A_z(\epsilon)}, $$
where $\gamma$ is the determinant of the flat boundary metric at $u\to 0$ and $S_{D7}^{\text{sub}}$ is the divergent subtracted action. Moreover,
$$\delta S_{D7}^{\text{sub}}=\int_{\epsilon} du\,\frac{\delta S_{D7}^{\text{sub}}}{\delta \partial_u A_z}\delta \partial_u A_z =-j \delta A_z(\epsilon),$$
where we have used the definition of $j$ in \eqref{betaDefi}. It immediately leads to $\langle J_z\rangle=-j$. For the details, see the appendix A of \cite{Hoyos:2011us}. 	
}. 
The equations \eqref{Q2}, \eqref{PPdefini} and \eqref{horLikePoint} give rise to the following relation for the current\footnote{From  \eqref{LegendreAction}, in fact, we find two different values for $j$ which lead to the current $\langle J_z\rangle=\pm \sigma E+\sigma_B B$. We see from \eqref{gaugeFields} that the information of electric field sign will not show up in the DBI and consequently in the D7-brane action, whereas the magnetic field (and axial chemical potential) sign is present in the CS action in \eqref{PPdefini}. The $\pm$ sign comes from the missing information by squaring $E$. For that reason we reject conventionally the negative sign in the current equation.},
\bea \label{OHM-CME}
\langle J_z\rangle= \sigma E+\sigma_B B,
\eea
where
\bea\label{Conductis}
\sigma=\frac{N_c N_f}{4\pi^{3/2}}\sqrt{\frac{4 B^2/\lambda}{\sqrt{4E^2/\lambda+\pi^2 T^4}}+\sqrt{4E^2/\lambda+\pi^2 T^4}} ,\quad \sigma_B=\frac{N_c N_f}{2\pi^2}\mu_5.
\eea
The $\sigma$ and $\sigma_B$ are studied previously in \cite{Karch:2007pd} and \cite{Hoyos:2011us}, respectively. Setting $\mu_5$($=\omega/2$) equal to zero, the CS action vanishes and consequently $\sigma_B=0$. Unlike $\sigma$, the magnetic conductivity $\sigma_B$ do not depend on the 't$\,$Hooft coupling. In fact, CME in the massless quark limit and strong coupling is similar to the calculations in the weak coupling. The term $\sigma_B$ is anomaly-induced transport coefficient and we know that the chiral anomaly is one-loop exact (for more discussions refer to \cite{Gursoy:2014ela}  and references therein). However, massive quarks break the chiral symmetry explicitly and we do expect that CME in strongly coupled differs from that in weak coupling. It can be seen that CME in the large $N_c$ and 't$\,$Hooft coupling receives stringy corrections for finite quark mass \cite{AliAkbari:2012if}.

\section{Chiral Equilibrium Instability of Quarkonium Mesons}\label{chiralInstSec}

The response of a QCD-like matter to the external electric and magnetic field at finite temperature and axial chemical potential has been studied in the previous section. We should note that the quarkonia are neutral degrees of freedom and we do not expect any interactions between them and the electromagnetic fields. It means that the electric current exists only in the phase of the melted mesons\footnote{We call the supersymmetric mesons in this study as quarkonium mesons, quarkonia or simply mesons interchangeably.}.

Let us first review the physical picture of the instability in the presence of electric field  \cite{Hashimoto:2013mua}. For a fixed quark mass, consider a system with no electric field initially and turn it on in a short interval $\delta t$ such that the final value of the electric field is strong enough to dissociate the quarks inside the mesons. In this case, the system is unstable and we expect that long enough time after turning on the field, the system reaches to the steady current $\vec{J}=\sigma \vec{E}$. The full time evolution of the system needs a time-dependent calculation in the gravity, however, the instability of equilibrium state just after applying the electric field can be studied by the imaginary part of the D7-brane action. Both approaches have been studied in  \cite{Hashimoto:2013mua}.

The electric current is also induced along the magnetic field in a plasma with non-zero axial chemical potential. Note that (as we mentioned in the introduction) there are some mesons in the supersymmetric QCD which are $U_{A}(1)$ charged. Now, assume the axial chemical potential of the system is below a critical value $\mu_{5,\text{cr}}$ and suddenly is increased above it. It leads to increasing the internal energy of the plasma and a phase transition happens. In this process, the electrically and chirally charged quarks are liberated from meson confinement, and eventually, CME leads to a steady current $\vec{J}=\sigma_B \vec{B}$ in the presence of external magnetic field.  The increasing of axial chemical potential may happen by a vacuum tunneling in the gluonic sector \cite{Kharzeev:2007jp} where the tunneling leads to a strong enough chiral imbalance in the system such that the axial chemical potential is raised above the critical value.

The time evolution of the induced current is not addressed in this study and we will only focus on the decay rate of the equilibrium state exactly after changing the axial chemical potential by studying the imaginary part of the D7-brane action. Moreover, we set the electric field equal to zero to suppress the Schwinger effect in this section. 

\subsection{Instability of Gapless Mesons}\label{gaplessSubSec}

In order to clarify the main idea of the paper, we investigate the (almost) gapless system in the large magnetic field in this subsection. The result for the small magnetic field is presented in the appendix \ref{appenSmall}.

In our holographic picture, it is not reasonable to speak about mesons in a system with massless quarks. In fact, at zero temperature and axial chemical potential and in the absence of quark mass scale, the system is conformal and there is no mass spectrum for mesons \cite{Kruczenski:2003be}. Here, we assume the quark mass is extremely small compare to the other physical quantities in the system such as temperature and axial chemical potential. Hence, one can set $\theta(u)\sim 0$ and the resulting D7-brane action is \eqref{LegendreAction}. The other assumption is that the magnetic field does not reach to its critical value where the meson melting transition disappears \cite{Albash:2007bk}. Above this critical value, the $\theta(u)\sim 0$ is not the preferred solution energetically and our calculations is not reliable anymore. This point has been studied in \cite{O'Bannon:2008bz} for the case of non-zero magnetic field and axial chemical potential.

Exactly after increasing the axial chemical potential suddenly, the electric current should be equal to zero.  Regarding this argument (and following \cite{Hashimoto:2013mua}), we set $j=0$ in the action \eqref{LegendreAction} and study  its possible imaginary part. After defining the following dimensionless quantities,
\bea \label{dimlessQuant}
y\equiv\frac{L}{u},\quad b\equiv (2\pi\alpha')\,B,\quad w\equiv L\,\omega,
\eea
the explicit form of the action in this limit is
\bea \label{SimpAction}
I = -\N\,L \int_{y_h}^{\infty} dy\,  \sqrt{y^6+y^2\,b^2-\frac{b^2 w^2}{q_h(y)}}.
\eea
where $I\equiv \left.\hat{S}_{D7}\right|_{\beta=0}$ and $q_h(y)=(1-y^4_h/y^4)$. Note that all dimensionful parameters are collected behind the integral.

The field theory as well as the gauge/gravity computations show that the Schwinger effect vanishes for the zero electric field even if the magnetic field is non-zero. But, interestingly, the appearance of the new term inside the square root related to the axial chemical potential is a sign of a new source for the equilibrium instability at the presence of $\mu_5$ and magnetic field and absence of the electric field.

In the zero temperature, the term inside the square root is negative at its minimum $y=0$. Therefore, the imaginary part of the action comes from the integration in the range $y\in [0,y_0]$ where $y_0$ is the root of the polynomial $y^6+y^2\,b^2-b^2 w^2$. The problem becomes more simple in the limit $b\gg 1$ where we can find $y_0$ and $\Im I$ analytically. In this limit, we should be careful about the critical value of the magnetic field mentioned at the beginning of this subsection. For example, repeating the \cite{O'Bannon:2008bz} procedure, we can find $b_{\text{cr}}\sim 13$ for $w=1$. Hence, in any case, the magnetic field should be always below an upper bound.

Assuming the polynomial $y^6+y^2 b^2-b^2 w^2$ has a finite root at $b\gg 1$, one can simply ignore $y^6$ and find $y_0=w$. Around this point, we consider the expansion $y_0=w\sum_{k=0}a_{2k}(w^2/b)^{2k}$ for small values of $w^2/b$. The coefficients $a_{2k}$ can be achieved order by order in this expansion,
\bea \label{rootExpan}
y_0=w\left[1-\frac{1}{2}\left(\frac{w^2}{b}\right)^2+\cdots\right]. 
\eea
We could go further in the expansion, but, comparing to the numerical calculations show that it does not improve the result for $\frac{w^2}{b}\lesssim 0.2$.

Until the root is approximately $y_0\sim w$, we have $0\leq y \lesssim w$ in the range of integration and we can expand $\sqrt{y^6+y^2 b^2-b^2 w^2}$ for large $b$. Using $y_0$ from above, we find the following expansion for the $\Im I$,
\bea \label{BigBImaginaryAnal}
\Im I=\frac{N_c N_f}{8\pi^2} \mu_5^2 B\left[\lambda^{1/2}-\frac{5}{(2\pi)^2}\lambda^{3/2}\left(\frac{\mu_5^2}{B}\right)^2+\cdots\right].
\eea

Unlike the holographic pictures of Schwinger effect, the above imaginary part depends on the 't$\,$Hooft coupling which might be an evidence for the difference of instability in the weak coupling limit. Moreover, the relation \eqref{BigBImaginaryAnal} is an expansion for small $w^2/b$. Nevertheless, after substituting physical dimensionful quantities, one finds that it is an expansion with respect to the 't$\,$Hooft coupling. It shows that this result cannot be calculated by the perturbative methods\footnote{In the next subsection, we re-write  $\Im I$ in terms of mesons mass and its binding energy. In this case, the 't$\,$Hooft coupling does not appear explicitly.}.

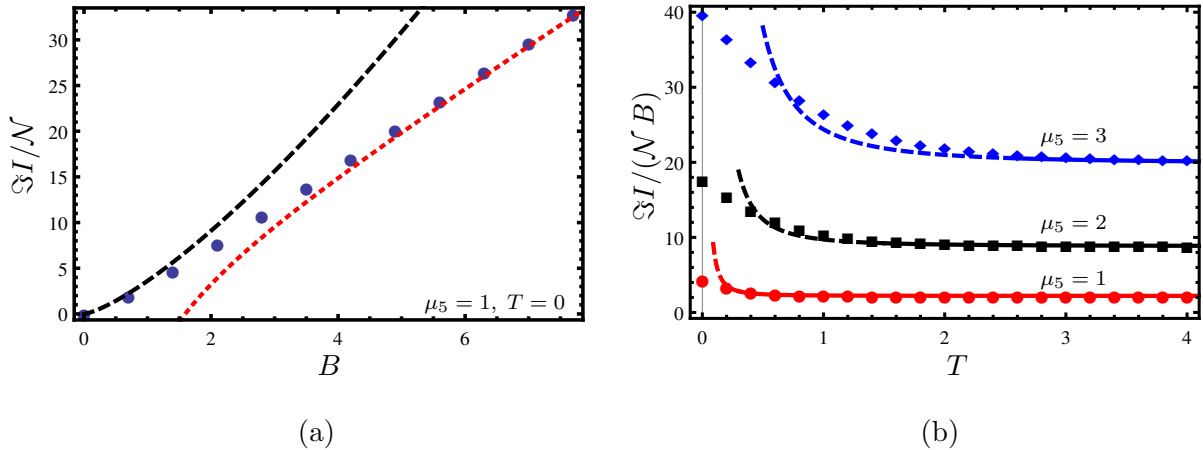
\begin{figure}[tb]
	\begin{center}
		\begin{tabular}{cc}
			
			\includegraphics[scale=0.80]{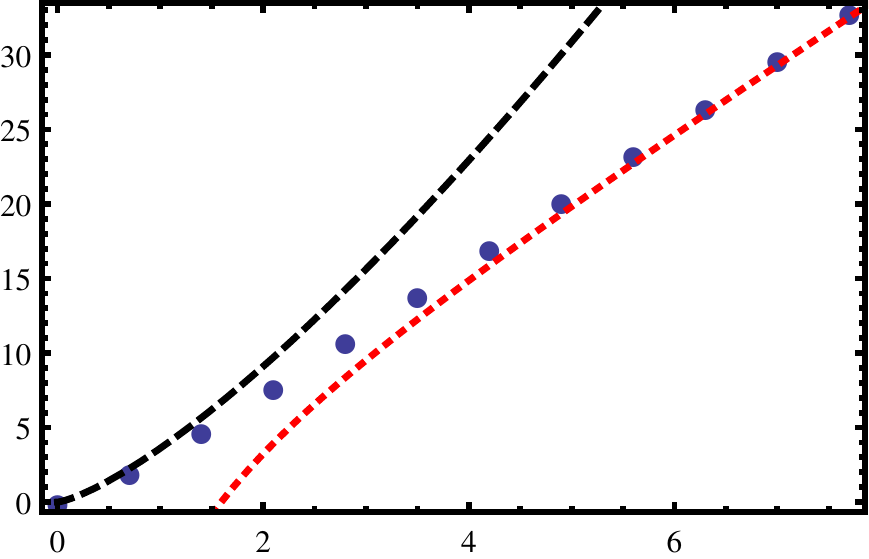}
			\put(-100,-10){$B$}
			\put(-60,15){\scriptsize{$\mu_5=1,\; T=0$}}
			\put(-215,70){\rotatebox[]{90}{$\Im I/\N $}}
			&
			\hspace{0.5cm}
			\includegraphics[scale=0.8]{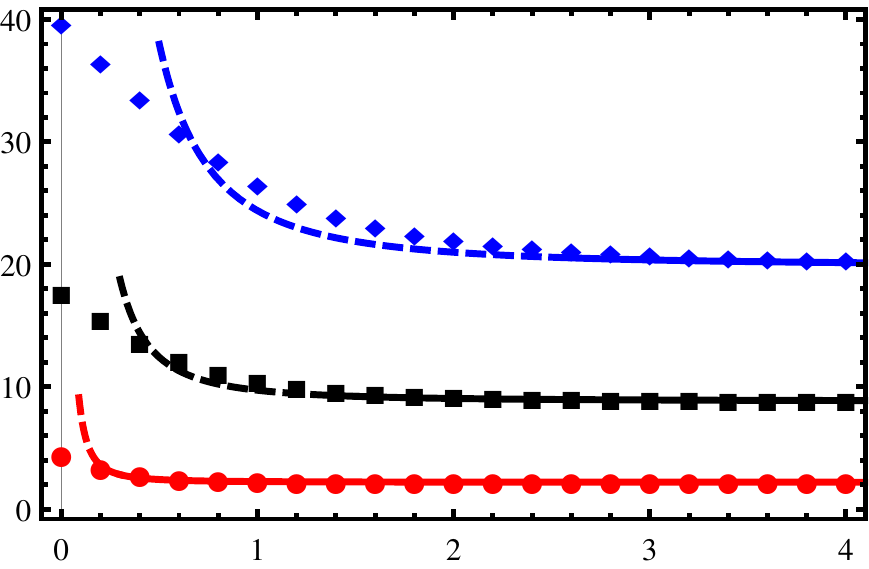}
			\put(-95,-10){$T$}
			\put(-225,145){\scriptsize{$\lambda=20$}}
			\put(-60,24){\scriptsize{$\mu_5=1$}}
			\put(-60,45){\scriptsize{$\mu_5=2$}}
			\put(-60,78){\scriptsize{$\mu_5=3$}}				
			\put(-213,70){\rotatebox[]{90}{$\Im I/(\N \,B)$}}\hspace{0.7cm}
			\vspace{0.5cm}
			\\
			(a) & (b)
		\end{tabular}
		\caption{(a) The imaginary part of the action at $T=0$. The black (dashed) curve is the analytical result for the small magnetic fields \eqref{LowBzeroT} and red (dotted) curve refers to the analytical result for the large  magnetic fields, \eqref{BigBImaginaryAnal}. The blue dots are obtained numerically. (b) Temperature dependence of $\Im I $ normalized to constant $\N$ and  magnetic field $B$ for three different values of axial chemical potential at large $B$.}
		\label{nonZerTemHighB}
	\end{center}
\end{figure}

In general, one can find $y_0$ and calculate the integration in \eqref{SimpAction} numerically. This has been done and the result is plotted in Fig.\,\ref{nonZerTemHighB}(a) for $\mu_5=1$ and different values of the magnetic field ($\lambda=20$). The numerical values of $\Im I$ normalized to the coefficient $\N$ are plotted by the blue dots in this figure\footnote{Here after in numerics, we use a unit in which $L=1$ and all dimensionful quantities such as magnetic field chemical potential \textit{etc.} also will be in this unit.}. Also the analytical results for two cases $b\ll 1$ and $b\gg 1$ are depicted by black dashed curve and red dotted curve, respectively. Both asymptotic analytical results \eqref{LowBzeroT} and \eqref{BigBImaginaryAnal} are in good agreement with numerical calculations.

Let us now study $\Im I$ in the large magnetic field and finite temperature.  In this limit, if the negative part of the term inside the square root is in a finite range of $y$ then the term $y^6$ can be ignored in \eqref{SimpAction}. As we will see, it is the case here. This limit brings the magnetic field $b$ out of the square root. If one rescales the integration parameter $y=w\,\tilde{y}$, the imaginary part of the simplified action turns into the following form,
\bea \label{finitTempAct}
\Im I = -\N\,L\, b\,w^2 \int_{\tilde{y}_h}^{\tilde{y}_0} d\tilde{y}\, \tilde{y} \sqrt{p_h(\tilde{y})},
\eea
where 
\bea
p_h(\tilde{y})=\frac{ \tilde{y}^2}{\tilde{y}^4-\tilde{y}_h^4}-1,
\eea
and $y_0$ is the root of $p_h(y)$. By inspection, one can find that for $\tilde{y}>\tilde{y}_h$ the function $p_h(\tilde{y})$ is monotonically decreasing and in the range $y\in[y_h,\infty)$, the root is as follows,
\bea \label{FinitTemRoot}
\tilde{y}_0=\left(\frac{1+\sqrt{1+4\tilde{y}_h^4}}{2}\right)^{1/2}.
\eea
It confirms the approximation of neglecting the term $y^6$ in \eqref{SimpAction}.

Note that $p_h(\tilde{y})$ goes to infinity in the limit $\tilde{y}\to \tilde{y}_h $. Let us introduce $\tilde{p}_h(\tilde{y})$ and demand that it has same infinite properties as $p_h(\tilde{y})$. Considering \eqref{FinitTemRoot}, it can be simply seen that the condition $\tilde{y}_h\gg 1$ (high temperature limit)  leads to $\tilde{y}_0=\frac{1}{4 \tilde{y}_h}+\tilde{y}_h$. In other words, large values of $\tilde{y}_h$  shrinks the integration range of $\Im I$ to a small interval with length $\frac{1}{4 \tilde{y}_h}$. Therefore, one can set $\tilde{y}=\epsilon+\tilde{y}_h$ and expand the function as follows,
\bea
p_h(\tilde{y})=\frac{1}{4 \epsilon\,\tilde{y}_h}+\text{finit term}+\mathcal{O}(\epsilon).
\eea
As a result, we choose $\tilde{p}_h(\tilde{y})=\frac{\tilde{y}_h}{4 (\tilde{y}-\tilde{y}_h)}+k$ and fix coefficient $k$ such that $\tilde{p}_h(\tilde{y}_0)=0$. The final result is the following,
\bea
\tilde{p}_h(\tilde{y})=\frac{1}{4\tilde{y}_h}\left(\frac{1}{ \tilde{y}-\tilde{y}_h}-\frac{1}{\tilde{ y}_0-\tilde{y}_h}\right).
\eea
It can be simply checked that the difference between $p_h(\tilde{y})$ and $\tilde{p}_h(\tilde{y})$ for $\tilde{y}_h>2$ ($T/\mu_5>4/\pi$) in the integration interval is less than $0.2\,\%$. The integration of approximated function can be performed and the result is \footnote{The following integration formula is used,
	\bea \label{integralForm}
	\int_{\alpha}^\beta dx\,x^n\,\sqrt{\frac{1}{x-\alpha}-\frac{1}{\beta-\alpha}}=\frac{\pi}{2} \alpha^n\sqrt{\beta-\alpha}\;\; 
	{}_2F_1\left(\frac{1}{2},-n,2,1-\frac{\beta}{\alpha}\right),
	\eea
	where ${}_2F_1$ is the hypergeometric function.
}
\bea \label{nonT-highB}
\Im I=\frac{N_c N_f}{16\pi^2}\lambda^{1/2} \mu_5^2 B\left[1+\left(\frac{1}{2\pi}\frac{\mu_5}{T}\right)^{2}+\mathcal{O}(\frac{\mu_5^4}{T^4})\right]+\mathcal{O}(\frac{\mu_5^4}{B^2}).
\eea

The interesting point of the above result is that it is not depend on $T$ in the first order of high temperature expansion. The other interesting point is that the above result defers by a factor of $1/2$ in comparison with the first term in the relation \eqref{BigBImaginaryAnal} at zero temperature. It shows that raising the temperature in the large magnetic field decreases the instability, meanwhile, the rate of decreasing becomes smaller at higher temperature  and finally $\Im I$ approaches to \eqref{nonT-highB}.

In more general case, we can study the equilibrium instability at high magnetic field numerically. The temperature dependence of $\Im I$ (normalized to constant $\N$ and large magnetic field $B$) for three different values of $\mu_5$ is shown in Fig.\,\ref{nonZerTemHighB}(b). It can be seen that the equilibrium instability decreases by increasing the temperature and approach to \eqref{nonT-highB}. The initial points in the numerical results are also in agreement with the first term \eqref{BigBImaginaryAnal}.

\subsection{Stable/Unstable Equilibrium State for Gapped Mesons}\label{gapped}

In this subsection, we extend our study to the massive quarks where the function $\theta(u)$ is not equal to zero anymore. One needs to calculate its functionality by solving the equation of motion. The asymptotic expansion of the solution is $\theta(u)= c_0 u+c_2 u^3+\cdots$ where $c_0$ is proportional to the quark mass $m$  and $c_2$ is proportional to the quark condensate $ \langle\bar{\psi}\psi\rangle$.

Before exploring the equilibrium instability of chiral mesons, let us mention some remarks on holographic CME \cite{Hoyos:2011us}. The  D7-brane embedding $\theta(u)$ (or equivalently $R(r)$ in the coordinate \eqref{OldCoord}) is achieved by the equation of motion calculated by varying  the action  \eqref{actionColli}. 
Due to the rotation of the branes, a worldvolume horizon might induce on the D7-brane which is given by
\bea\label{WVHorizon}
\theta_{\text{wv}}(u_*)=\arcsin\sqrt{\frac{1}{w^2u_*^2}\left(1-\frac{u_*^4}{u_h^4}\right)}.
\eea
At the finite temperature and axial chemical potential, there are two possible horizons, background horizon of AdS-Schwarzschild and worldvolume horizon. Accordingly, there are three different embeddings for D7-branes which depend on the induction of background or worldvolume horizons on them \cite{Hoyos:2011us,AliAkbari:2012if,Ali-Akbari:2014nua}.
The CME is non-zero for embeddings that intercept the above locus.  Using this picture, one can relate the current to the quark mass as follows. By increasing the quark mass, there is a critical mass (critical embedding) where the D7-brane do not cross the locus \eqref{WVHorizon} which means CME is zero for $m>m_{\text{cr}}$  (see figure 2 of Ref.\cite{Hoyos:2011us}). In the field theory picture, for small mass, there are some melted mesons or equivalently quarks that move along the magnetic field via CME. However, for larger mass, there are neutral bound state of quarks in the system, therefore, there is no current.

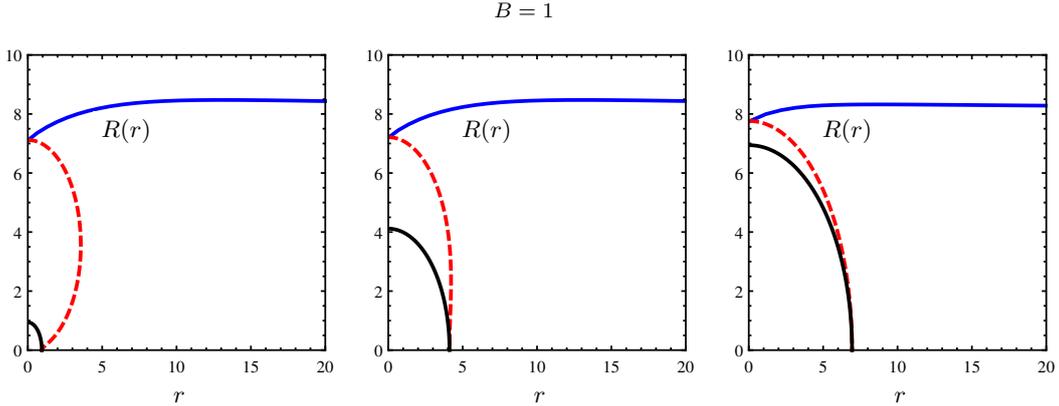
\begin{figure}[tb]
	\begin{center}
		\begin{tabular}{ccc}
			\includegraphics[scale=0.7]{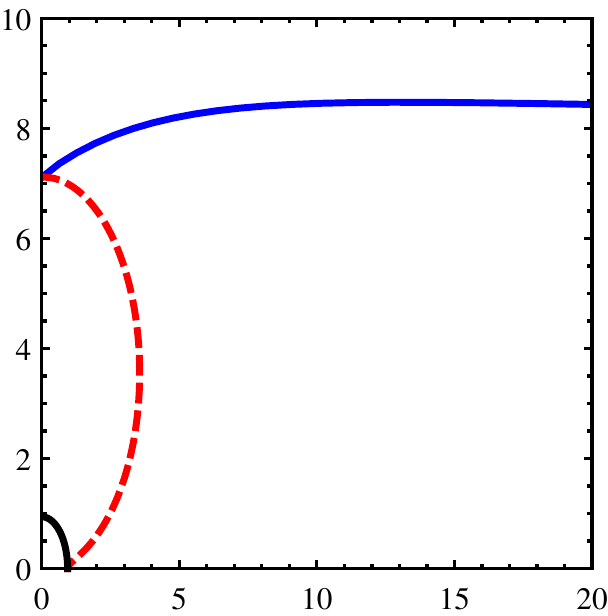}
			\put(60,135){\scriptsize{$B=1$}}
			\put(-90,90){\colorbox{white}{\footnotesize{$R(r)$}}}
			\put(-63,-10){\colorbox{white}{\footnotesize{$r$}}}
			&
			\includegraphics[scale=0.7]{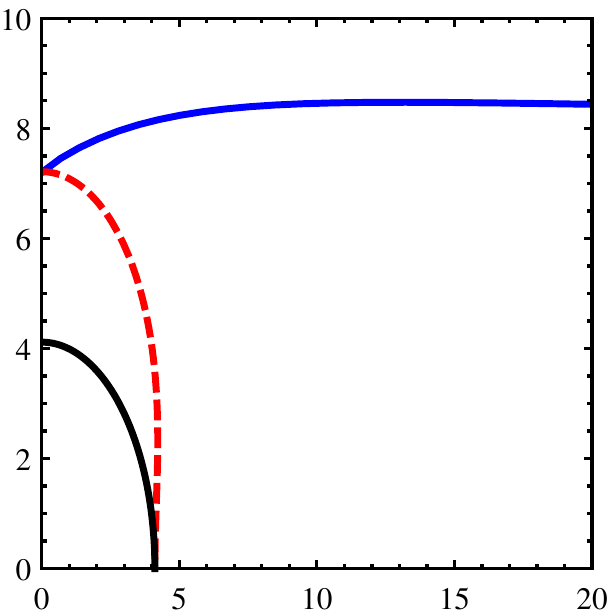}
			\put(-90,90){\colorbox{white}{\footnotesize{$R(r)$}}}
			\put(-63,-10){\colorbox{white}{\footnotesize{$r$}}}
			&
			\includegraphics[scale=0.7]{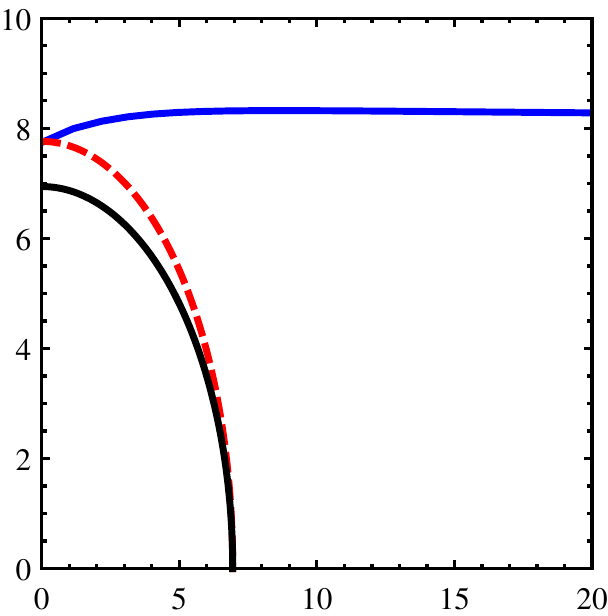}
			\put(-90,90){\colorbox{white}{\footnotesize{$R(r)$}}}
			\put(-63,-10){\colorbox{white}{\footnotesize{$r$}}}			
		\end{tabular}
		\caption{The blue curves are the D7-brane embedding $R(r)$ for $T=0.3,\,\mu_5=3.56$ (left), $T=1.31,\,\mu_5=3.41$ (middle), $T=2.21,\,\mu_5=2.33$ (right). The asymptotic value of $R(r)$ corresponds to the mass $m=5.81$ for $\lambda=20$. Red (dashed) curves indicate the worldvolume horizons and black curves are black hole horizons.
		}
		\label{ThreeEmbedding}
	\end{center}
\end{figure}

\begin{figure}[tb]
	\begin{center}
		\begin{tabular}{cc}
			\includegraphics[scale=0.73]{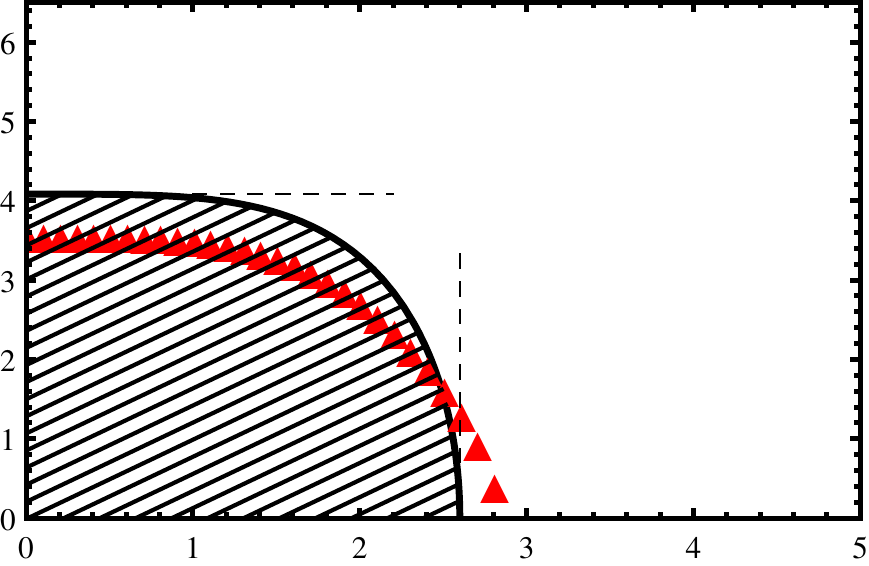}
			\put(-95,-10){$T$}
			\put(2,145){\scriptsize{$\lambda=20$}}
			\put(-120,80){\scriptsize{$\mu_{5,\text{cr}}$}}
			\put(-85,55){\rotatebox[]{-90}{\scriptsize{$T_{\text{cr}}$}}}				
			\put(-195,60){\rotatebox[]{90}{$\mu_5$}}\hspace{0.7cm}
			\put(-170,27){\colorbox{white}{\small{Stable Region}}}	
			\put(-90,90){\colorbox{white}{\small{Unstable Region}}}	
			&
			\hspace{0.4cm}
			\includegraphics[scale=0.75]{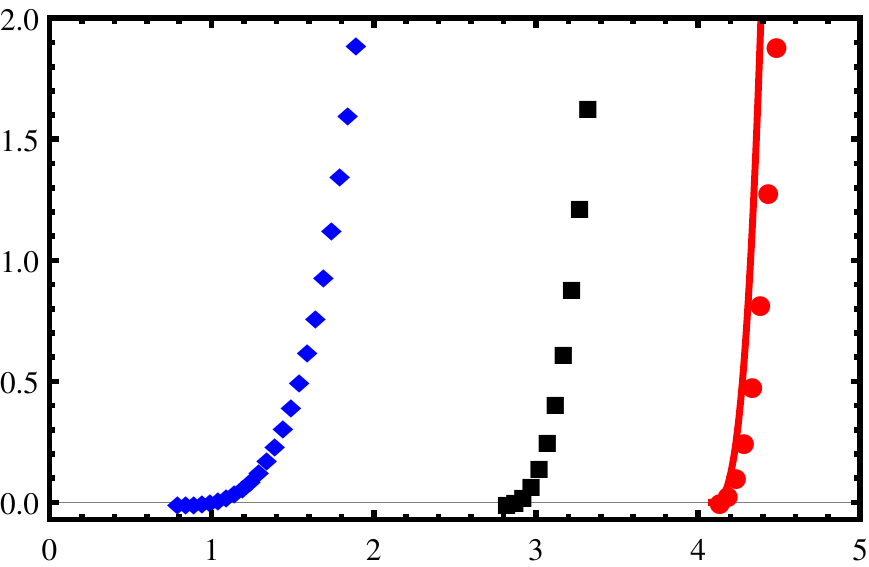}
			\put(-105,-10){$\mu_5$}
			\put(-125,80){\rotatebox[]{84}{\scriptsize{$T=2.58$}}}
			\put(-73,80){\rotatebox[]{84}{\scriptsize{$T=2.23$}}}
			\put(-33,80){\rotatebox[]{86}{\scriptsize{$T=0$}}}				
			\put(-200,70){\rotatebox[]{90}{$\Im I/\N$}}
			\vspace{0.5cm}
			\\
			(a) & (b)
		\end{tabular}
		\caption{(a) The phase diagram of stable/melted mesons (red triangles) and stable/unstable equilibrium state (black curve). (b) The $\mu_5$ dependence of $\Im I/\N$ for three different temperatures. The dots refer to numerical calculations and red curve corresponds to analytical result for $T=0$. The $m=5.81$ has been considered for quark mass.}
		\label{PhaseDiagI}
	\end{center}
\end{figure}
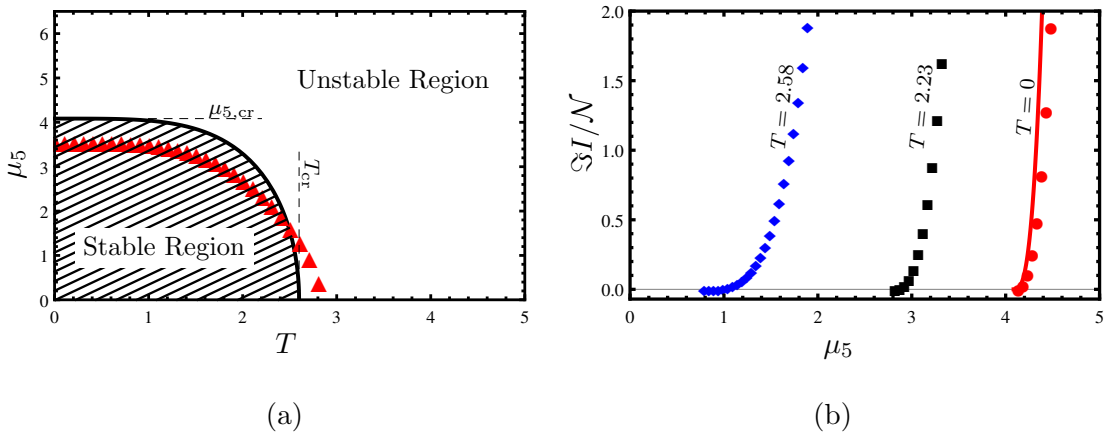

According to the above picture, we can plot a stable/melted phase diagram for a given quark mass. One should fix the mass and check for what values of $T$ and $\mu_5$ the current becomes non-zero. We have shown three different critical embeddings for same quark mass in Fig.\,\ref{ThreeEmbedding}. The red dashed curve indicates worldvolume horizon and black curve depicts the background horizon and the blue curve is the critical embedding. All the curves are in \eqref{OldCoord} coordinate. Different values of $T$ and $\mu_5$ for same quark mass $c_0\simeq 8.17$ are shown by red triangles in Fig.\,\ref{PhaseDiagI}(a). Here the value of magnetic field is chosen $B=1$ which is much lower than its critical value. For $T=1.5$ and $\mu_5=3.5$ the critical magnetic field is $B_{\text{cr}}\sim 500$. One can check that the phase diagram in Fig.\,\ref{PhaseDiagI}(a) does not change drastically by choosing $B=10$ and $30$.

Now we would like to find out the decay rate of a system at temperature $T$ when the value of $\mu_5$ suddenly increases. Basically, we could assume that the quarkonia at given $T$, $\mu_5=0$ and external magnetic field is in the equilibrium state. It leads to a numerical function for $\theta(u)$. Using it, we calculate the terms inside the square root in the action and study its value for non-zero $\mu_5$ and investigate whether it becomes imaginary. Note that the existence of non-zero $\mu_5$ deforms the D7-brane shape, but, if it happens very rapidly then we can assume that the initial shape of the brane does not deviate from $\theta(u)$. The same logic is used for studying the Schwinger effect in the gapped systems in the presence of electric field \cite{Hashimoto:2013mua}. In \cite{Hashimoto:2013mua}, the initial embedding is the supersymmetric solution for D7-brane because the temperate is assumed to be zero initially. In $T=E=B=\mu_5=0$, the supersymmetric solution for D7-brane embedding is
\bea \label{superSolu}
\theta(u)=\arcsin\left[\frac{2\pi\,L^2\,m}{\sqrt{\lambda}}u\right],
\eea
where $m=c_0/(2\pi\alpha')$. This solution in the coordinate \eqref{OldCoord} is simply $R(r)=c_0$. Following this logic, we assume at finite temperature and zero chemical potential the solution $\theta(u)$ is approximately what is mentioned in \eqref{superSolu}. This is a good approximation in the following conditions: $i$) The magnetic field is much smaller than its critical value, or $ii$) The increase in the axial chemical potential and the magnetic field  happens simultaneously. In fact, this approximation becomes exact if all variables turn on at the same time, however, we assume the first option, $B\ll B_{\text{cr}}$.

For the massive quarks, the parameters $\QQ_1$, $\QQ_2$ and $\PP_1$ in \eqref{Q1}, \eqref{Q2} and \eqref{PPdefini} change to the following forms,
\bea
\QQ_1&=&\frac{L^{10}}{u^{10}}\left(1+(2\pi\alpha')^2B^2\frac{u^4}{L^4}\right)\left(1-(2\pi\alpha')^2\frac{m^2\,u^2}{L^4}\right)^3,\nn\\
\QQ_2&=& \frac{L^6}{u^6}b_h(u)\left(1+(2\pi\alpha')^2B^2\frac{u^4}{L^4}\right)\left(1-(2\pi\alpha')^2\frac{m^2\,\omega^2\,u^4}{L^4\,b_h(u)}\right)\left(1-(2\pi\alpha')^2\frac{m^2\,u^2}{L^4}\right)^3(2\pi\alpha')^2,\nn\\
\PP_1&=& L\,(2\pi\alpha')^2\,B\,\omega \,\left(1-(2\pi\alpha')^2\frac{m^2\,u^2}{L^4}\right)^2.
\eea
By defining $\eta\equiv \frac{2\pi\alpha' \, m}{L}=\frac{2\pi}{\sqrt{\lambda}}m\,L$
and using redefinitions \eqref{dimlessQuant}, the action can be rewritten as
\begin{equation}\label{MoreGenAction01}
\begin{split}
I=-\N\,L\int_{y_{\min}}^{\infty}\frac{dy}{y^3}\sqrt{(y^2-\eta^2)^3\left[y^6+y^2\left(b^2-\frac{w^2\,\eta^2}{q_h(y)}\right)-\frac{b^2\,w^2}{q_h(y)}\right]},
\end{split}
\end{equation}
where $y_{\min}=\max\{\eta,y_h\}$.

We would like to find a stable region in the $T$, $B$ and $\mu_5$ space. First assume $\eta\geq y_h$ which means $(y^2-\eta^2)^3\geq 0$. In \eqref{MoreGenAction01}, the term in the square bracket should be positive in the range  $\eta\leq y<\infty$ to avoid the negative values inside the square root. It happens if we choose $|w|>\sqrt{\eta^4-y_h^4}/\eta$ or equivalently, 
\bea\label{stableRegion}
|\mu_5|>\sqrt{\frac{\pi^2 m^2}{\lambda}-\frac{\pi^2\lambda\,T^4}{16m^2}}.
\eea
In the region $\eta<y_h$, however, there is always a window that the action \eqref{MoreGenAction01} becomes imaginary. In this region, $(y^2-\eta^2)$ has always a positive value, but the term inside the square bracket goes to $-\infty$ at $y\to y_h$ and $+\infty$ at $y\to \infty$ which means it changes its sign somewhere in the middle. The relation \eqref{stableRegion} shows the stable/unstable region for the given $T$ and $\mu_5$. Note that in this relation $B$ is absent. It is compatible with the previous remark that the magnetic field has not active role in the quarkonia instability. However, we should mention that the approximation \eqref{superSolu} leads to the magnetic field independence in the stable/unstable regions. We expect the value of magnetic field (well below its critical value) changes mildly the stable/unstable region in a more accurate study. There is a critical value for the temperature where above that, the system becomes unstable for any values of axial chemical potential. Similarly a critical value for $\mu_5$ is obtained,
\bea
T_{\text{cr}}=\frac{2\,m}{\sqrt{\lambda}},\qquad \mu_{5,\text{cr}}=\frac{\pi\,m}{\sqrt{\lambda}}.
\eea

The stable region \eqref{stableRegion} is the region where the supersymmetric solution \eqref{superSolu} for a fixed mass does not intercept the worldvolume horizon \eqref{WVHorizon} for different values of $T$ and $\mu_5$. In other words, one can find \eqref{stableRegion} by using the supersymmetric solution and the method we have found the stable/melted meson phases. This is the shaded region in Fig.\,\ref{PhaseDiagI}(a). The difference between stable region (shaded region) and stable meson phase (the region below the red triangles) is that the later is achieved in the equilibrium state where $T$ and $\mu_5$ are assumed to be fixed, while the former is calculated for the case $\mu_5$ is suddenly increases from zero. In the gravity side, it can be understood by the difference between two solutions of $\theta(u)$ in two cases.

Now we will study the decay of the equilibrium state when $\mu_5$ increases suddenly such that the system goes into the unstable region in $T$-$\mu_5$ space. In the limit $y_h=0$ and $b\gg 1$, it is possible to find an analytical result. The imaginary part of the action \eqref{MoreGenAction01} in this limit is
\bea
\Im I =\N\,L\,b \int_{\eta}^{w}\frac{dy}{y^3}\sqrt{(y^2-\eta^2)^3(w^2-y^2)}=\N\,L\,b \frac{\pi (w-\eta)^3}{4w},
\eea
which can be written in terms of the field theory quantities as,
\bea \label{GappedAnal01}
\Im I = \frac{N_c\,N_f}{8\pi^2}\sqrt{\lambda}\mu_5^2\,B\left[1-\frac{\pi}{\sqrt{\lambda}}(\frac{m}{\mu_5})\right]^3\qquad \mu_5\geq \mu_{5,\text{cr}}.
\eea
Interestingly, only the term in the square bracket is added compare to the gapless system (see \eqref{BigBImaginaryAnal}). Recall that adding mass in the Schwinger effect \eqref{SchwingerOrigin} leads to a summation of exponentials  which is interpreted as a sum over instantons.

The other feature appears if we write $\Im I$ in terms of the quarkonium mass $M_q$ and its binding energy $E_b$. In this case, the 't$\,$Hooft coupling does not appear explicitly in the equation. The mass of the lightest meson in the model under consideration is given by \cite{Kruczenski:2003be}
$$M_q=4\pi\sqrt{2}\frac{m}{\sqrt{\lambda}}=4\sqrt{2}\mu_{5,\text{cr}}.$$
According to the above, the meson mass is much smaller than its quark constituents. Consequently, the binding energy is very large in this model and is given by $E_b\sim 2\,m$. Using these facts, we have
\bea \label{bindingMesonInst}
\Im I = \frac{N_c\,N_f}{8\pi^2}\mu_5^2\,B\left[1-\frac{M_q}{4\sqrt{2}\,\mu_5}\right]^3\left(\frac{2\pi\sqrt{2}\,E_b}{M_q}\right)\qquad \mu_5\geq \frac{M_q}{4\sqrt{2}}.
\eea
This is true for large number of colors and large 't$\,$Hooft coupling of $\N=4$ SYM theory and what we could ask is whether it is true for QCD. Unlike the mesons in our model, the binding energy is not large in QCD. As an example the $J/\Psi$ meson binding energy is around $\sim 0.4\, \text{GeV}$ which is one order smaller than its mass and constituent quarks. In any case, according to the above result, the system is more unstable for mesons with larger binding energies.

In more general cases, a numerical analysis is needed to find $\Im I$. It can be found in Fig.\,\ref{PhaseDiagI}(b) for $T\simeq0$ (red), $T\simeq 2.23$ (black) and $T\simeq 2.58$ (blue). The quark mass is chosen $m\simeq 5.81$ and $B\simeq 35.6$. It is seen that the $\Im I$ is non-zero when the $\mu_5 \geq \mu_{5,\text{cr}}$. The red curve corresponds to the analytical result for instability \eqref{GappedAnal01}. There is a good agreement between the analytical and the numerical results for $T=0$.

\section{Melted Meson Equilibrium Point and its Instability}\label{EBMUT}

In this section, we again study a gapless system, but this time, in the presence of electric and magnetic field at the same time. One can imagine a case that both Ohm's law and CME produce current in opposite directions such that the net current is equal to zero. In this case, there is a sub-space in $(E,B,T,\mu_5)$ space in which the system is always in equilibrium\footnote{We should be careful in considering the state in equilibrium. In fact, Ohm's law is a dissipative effect and CME is non-dissipative. Here equilibrium state is referred to a state with no current. }. In order to study the decay rate to a non-equilibrium state, we slightly move away from this sub-space and then study the instability.

 Choosing $e\equiv (2\pi\alpha')\,E$, the D7-brane action is written as
\bea \label{SimpActionElect}
I = -\N\,L \int_{y_h}^{\infty} dy\,  \sqrt{p_e(y)\,p_b(y)},
\eea
where
\bea
p_e(y)=1- \frac{e^2}{y^4 q_h(y)},\quad p_b(y)=y^6+y^2\,b^2-\frac{b^2 w^2}{q_h(y)}.
\eea
Let us define $y_-=\text{min}\{y_e,y_b\}$ and $y_+=\text{max}\{y_e,y_b\}$ where $y_e$ and $y_b$ are roots of $p_e(y)$ and $p_b(y)$, respectively. Then the imaginary part of the action comes from integration of $\sqrt{p_e(y)\,p_b(y)}$ in the range $y_-<y<y_+$ . When $y_-\to y_+$ the $\Im I=0$ so the system is stable. This happens in a special point where $p_e(y)$ and $p_b(y)$ have common roots. Such a point exists if the equation $p_b(y_e)=0$ is satisfied for at least one point in $(E,B,T,\mu_5)$ space. Using \eqref{Conductis}, this equation reduces to\footnote{Similar to the current in equilibrium \eqref{OHM-CME}, there is a sign ambiguity. It happens because sign of electric field is missed by squaring $E$ in DBI action. However, the sign of $B$ and $\mu_5$ is stored in CS action. We choose minus sign compatible with \eqref{OHM-CME}.}
\bea \label{StablePoint}
\sigma_B B =- \sigma E,
\eea
where $\langle J_z\rangle=0$. Here, we will try to find $\Im I$ when we slightly violate the constraint $\sigma_B B =- \sigma E$.

In general, numerical calculations are needed to find the $\Im I$. However, we can simplify it in some special limits and find analytical results. In the limit $B\to +\infty$, the electrical conductivity $\sigma$ is 
\bea
\sigma=\frac{N_c N_f}{2\pi^2}\left[\frac{4E^2}{\pi^2\,\lambda}+T^4\right]^{-1/4}\frac{|B|}{\sqrt{\lambda}}. 
\eea
Now the equation \eqref{OHM-CME} can be written as
\bea\label{OHM-CME-HighB}
\langle J_z\rangle=\frac{N_cN_f}{2\pi^2}\left(\mu_5+\tilde{\mu}_5\right)B+\mathcal{O}(\frac{1}{B}),
\eea
in the large magnetic field where
\begin{equation}\label{muTilde}
\begin{split}
\tilde{\mu}_5&\equiv  \frac{E}{\sqrt{\lambda}}\left[\frac{4}{\pi^2\,\lambda}E^2+T^4\right]^{-1/4}.\\
\end{split}
\end{equation}
Regarding this equation, $\tilde{\mu}_5$ mimics the axial chemical potential in large magnetic field and the equation \eqref{StablePoint} reduces to $\mu_5=- \tilde{\mu}_5$.

\begin{figure}
	\begin{center}
		\includegraphics[scale=0.80]{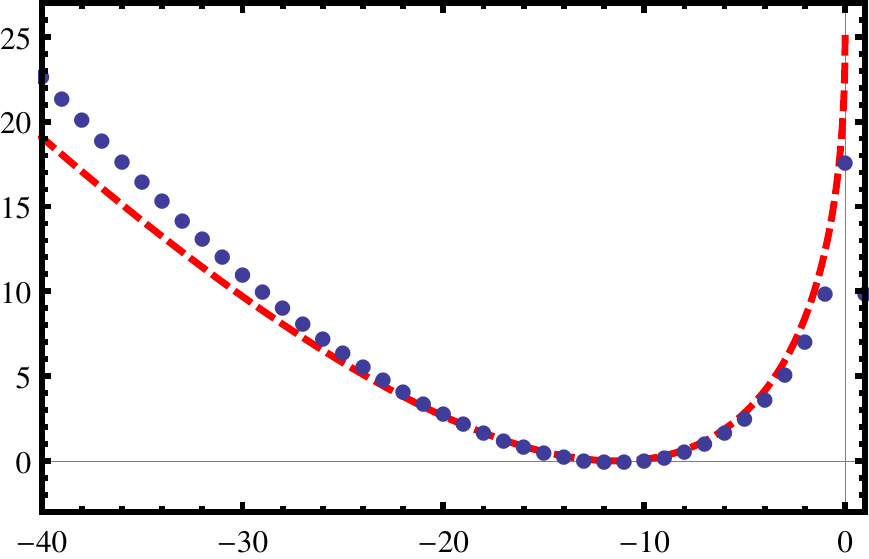}
		\put(-100,-10){$E$}
		\put(-95,14.5){\scriptsize{$\mu_5=2,\; \lambda=20,\; T=0$}}
		\put(-215,70){\rotatebox[]{90}{$\Im I/(\N\,B) $}}
		\caption{The electric field dependence of $\Im I/(\N B)$ for zero temperature and large magnetic. The blue dotes refer to numerical analysis and red (dashed) curve corresponds to analytical result.}\label{StablePoints}
	\end{center}
\end{figure}
Let us calculate $\Im I$ for a large magnetic field and zero temperature in a small interval with radius $\epsilon\to0$ around stable region $\mu_5=- \tilde{\mu}_5$.  In this limit, we have
\bea
\Im I =\N\, L\, b \int_{y_-}^{y_+} dy\,\sqrt{(e^2/y^4-1)(y^2-w^2)}
\eea
where $y_-=\text{min}\{\sqrt{|e|},|w|\}$ and $y_+=\text{max}\{\sqrt{|e|},|w|\}$ and  $y_+= y_-+\epsilon$. The term inside the square root can be factorized in the following form
\bea
\left[\frac{(\sqrt{|e|}+y)(|e|+y^2)(|w|+y))}{y^4}\right]\left[\frac{(|w|-\sqrt{|e|})^2}{4}-\left(y-\frac{|w|+\sqrt{|e|}}{2}\right)^2\right].
\eea
Inside the interval of integration we have $y\sim \sqrt{|e|}\sim |w|$. It indicates that the leading term inside the first bracket in the above relation does not depend on $\epsilon$ for finite $\sqrt{|e|}$ and $|w|$. However, the leading term in the second bracket is in the order of $\epsilon$. We approximate the first bracket by evaluating it around $(y_++y_-)/2$. One can check that the value of the first bracket in this approximation is equal $8+4\epsilon/|w|$. The integration of the second bracket inside the square root is nothing but the area of a semi-circle with radios $(y_+-y_-)/2$.  The final result is the following\footnote{There is a sign difference if $\sqrt{e} \gtrsim w$ or $\sqrt{e} \lesssim w$ that we absorb it into the sign ambiguity of $\Im I$.},
\bea \label{PerturbFromStable}
\Im I =\frac{N_c N_f}{4\pi^2}\sqrt{\frac{\lambda}{2}}\,B\left(|\mu_5|-|\tilde{\mu}_5|\right)^2+\cdots.
\eea
According to the above result, the leading term of the instability is quadratically increasing when we move away from the stable sub-space in large magnetic field and zero temperature.

In Fig.\,\ref{StablePoints}, the numerical calculations are compared with analytical results for $\mu_5=2 $. The stable point occurs at $E=-8\sqrt{\lambda}/\pi$. The red dashed curve in this figure is analytical result \eqref{PerturbFromStable} and blue dots are obtained by numerical calculations. It can be seen that two results have good agreement around the stable point $|\tilde{\mu}_5|=|\mu_5|$. Note that at $E\to 0$ the leading term inside the first square bracket is not large compare to $\epsilon$ anymore and the analytical result is not valid. The $\Im I$ has been already calculated in \eqref{BigBImaginaryAnal} for the case $E=0$ and it shows correct value in comparison with numerical solutions at this point.

As a final remark, let us mention that by setting $w=y_h=0$ in \eqref{SimpAction} the $\Im I$ diverges (for details please refer to \cite{Hashimoto:2014dza}). Interestingly, the existence of $\mu_5$ in our study leads to disappearance of this divergence even at the zero temperature.

\section{Summary}\label{summar}

Using AdS/CFT, equilibrium instability of chiral quarkonia  in a plasma in the presence of external electric and magnetic field and at finite axial chemical potential has been studied. We have found that despite the system is stable when there exist magnetic field only, in the case that it is accompanied by axial chemical potential the system may become unstable.  This instability has been investigated via imaginary part of the D7-branes action.

The stable/unstable phase transition and the instability of mesons (see Fig.\,\ref{PhaseDiagI}) have been investigated in the following picture. Consider a stable system of mesons in a plasma at finite temperature such that its axial chemical potential is much smaller than a critical value $\mu_{5,\text{cr}}$ (\textit{e.g.} $\mu_5=0$ initially). Also, a non-zero external magnetic field exists in the background. Recall that quarkonia in the system are charged with respect to $U_A(1)$ but they are not electrically charged. For that reason, they can not couple to magnetic field and no current can be induced by CME. However, by assuming an external mechanism (a vacuum tunneling for example) a sudden change in the axial chemical potential occurs such that $\mu_5$ goes beyond the critical value  $\mu_{5,\text{cr}}$. Then the system becomes unstable and constituent quarks of the quarkonium are liberated. In this case, the new degrees of freedom are electrically and axially charged and CME produces the current. It means that the initial equilibrium decays to another state which is ultimately a steady state with constant current. For quarkonia with mass $M_q$ and binding energy $E_b$ the $\Im I$ at zero temperature and large magnetic field has been presented in \eqref{bindingMesonInst}. Also the numerical study for more general cases have been depicted in Fig.\,\ref{PhaseDiagI}(b).

Finally, the massless mesons have been investigated in a system that two Ohm's law and  CME cancel their effects. These are two different mechanisms in producing current. Basically, it happens for specific choices of $E$, $B$ and $\mu_5$ (and $T$) such that $\sigma_B B =- \sigma E$. Any deviation from this sub-space makes the equilibrium state unstable. We have studied $\Im I$ when the constraint $\sigma_B B =- \sigma E$ is slightly violated.

\acknowledgments

We would like to thank M.\;Ali-Akbari,  H.\;Ebrahim, S.\;Fayazbakhsh, A.\;E.\;Mosaffa,  H.\;Arfaei,   A.\;Davodi, N.\;Abbasi, Y.\;Ayazi,  D.\;Allahbakhshi for useful discussions. We are also grateful to D.\;Kaviani for helpful and fruitful comments.


\appendix

\section{More About Instability of Gapless Systems}
\subsection{Small Magnetic Field}\label{appenSmall}

Recall the action \eqref{SimpAction}. At zero temperature, the term inside the square root is  $y^6+y^2 b^2-b^2 w^2$ which is negative in the range $0\leq y < y_0$. In the limit $b\to 0$, we can easily ignore the term $b^2 y^2$ and immediately find the root $y_0=(b\,w)^{1/3}$. Now obtaining the imaginary part of the action \eqref{SimpAction} is straightforward,
\begin{equation} \label{LowBzeroT}
\begin{split} 
\Im\,I&=\N\,L \int_{0}^{(b\,w)^{1/3}}dy\sqrt{b^2 w^2-y^6},\\
&= \frac{N_c N_f}{3}\mathcal{B}(\frac{1}{6},\frac{3}{2})\left[2\lambda \left(\frac{ B\,\mu_5}{4\pi^{2}}\right)^4\right]^{1/3},
\end{split}
\end{equation}
where $\mathcal{B}(1/6,3/2)$ is the beta special function.

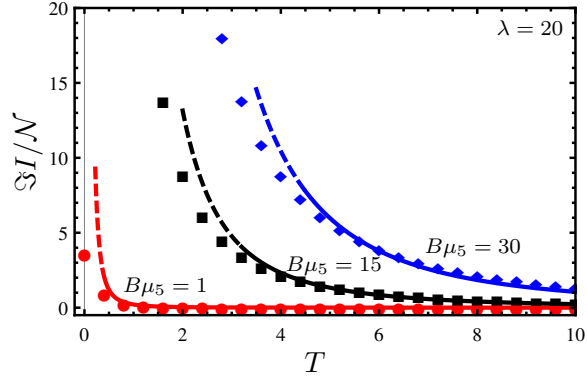
\begin{figure}[tb]
	\begin{center}
		\begin{tabular}{cc}
			\includegraphics[scale=0.8]{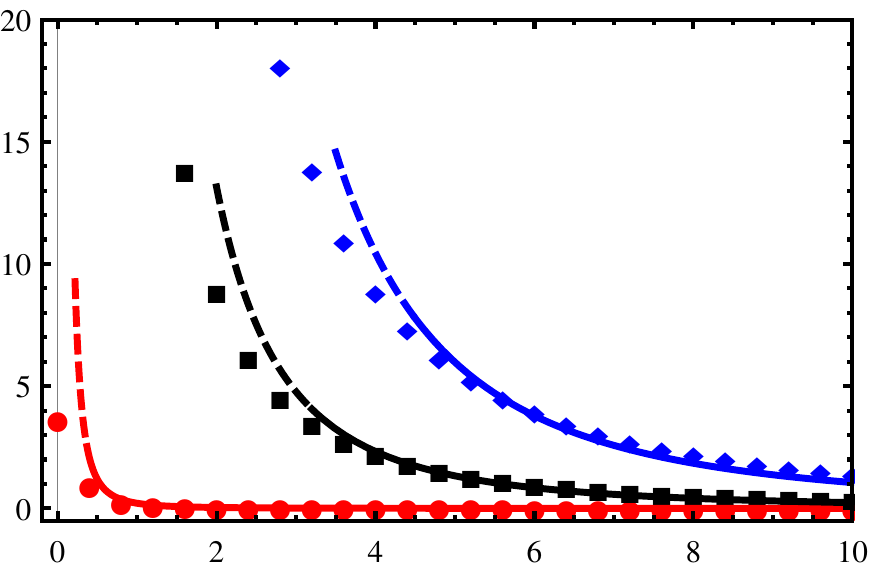}
			\put(-105,-10){$T$}
			\put(-173,20){\scriptsize{$B\mu_5=1$}}
			\put(-112,28){\scriptsize{$B\mu_5=15$}}
			\put(-60,35){\scriptsize{$B\mu_5=30$}}	
			\put(-33,117){\scriptsize{$\lambda=20$}}		
			\put(-214,70){\rotatebox[]{90}{$\Im I/\N$}}
			\vspace{0.5cm}
		\end{tabular}
		\caption{ Temperature dependence of $\Im I $ normalized to constant $\N$ for three different values of $B\,\mu_5$ at small $B$. Dotes refer to numerical calculations and curves refer to analytical results.}
		\label{nonZerTemHighBApp}
	\end{center}
\end{figure}

For the finite temperature case, we do as follows. By sending the magnetic field to zero and defining $y=(b\,w)^{1/3}\tilde{y}$ the imaginary part of the action \eqref{SimpAction} can be written as
\bea
\Im I = -\N\,L\, (b\,w)^{4/3} \int_{\tilde{y}_h}^{\tilde{y}_0} d\tilde{y}\, \tilde{y}^2 \sqrt{p(\tilde{y})},
\eea
where
\bea
p(\tilde{y})=\frac{1}{\tilde{y}^4-\tilde{y}_h^4}-\tilde{y}^2.
\eea
Similar to the case $b\gg w^2$ at zero temperature, the root of $p$ can be found perturbatively as the following,
\bea
\tilde{y}_0=\tilde{y}_h\left(1+\frac{1}{4 \tilde{y}_h^5}-\frac{7}{32\tilde{y}_h^{12}}+\cdots\right)
\eea
At high temperature, we will use the same strategy as subsection \ref{gaplessSubSec} and replace the function $p(\tilde{y})$ by $\tilde{p}(\tilde{y})$ where the asymptotic behavior of two functions at $\tilde{y}\to\tilde{y}_h$ are same and $\tilde{p}(\tilde{y}_0)=0$. The resulting function is
\bea
\tilde{p}(\tilde{y})=\frac{1}{4\tilde{y}_h^3}\left(\frac{1}{ \tilde{y}-\tilde{y}_h}-\frac{1}{\tilde{ y}_0-\tilde{y}_h}\right),
\eea
which is much easier to integrate analytically. The integration formula \eqref{integralForm}  leads to the analytical result for $\Im I$,
\bea \label{LowBNonZeroT}
\Im I =\frac{N_c N_f}{2^{8/3}\pi^{10/3}\lambda^{1/12}}(B\,\mu_5)^{4/3}\left(\frac{B\,\mu_5}{T^3}\right)^{5/6}+\cdots.
\eea
Several features of the above result can be considered. First, the 't$\,$Hooft coupling is appeared with negative power which means in contrast to the other limits, the effect vanishes at $\lambda\gg 1$. Second feature is that the instability disappears at high temperature limit.

The numerical results for arbitrary temperature, small magnetic field and three different values of $B\,\mu_5$ are presented in Fig.\,\ref{nonZerTemHighBApp}. For large enough values of $T$, the numerical results and the analytical expansion \eqref{LowBNonZeroT} are in agreement. In addition, the numerical values  at zero temperature are compatible with equation \eqref{LowBzeroT}.

\subsection{General Numerical Study}

\begin{figure}[tb]
	\begin{center}
		\begin{tabular}{cc}
			\includegraphics[scale=0.8]{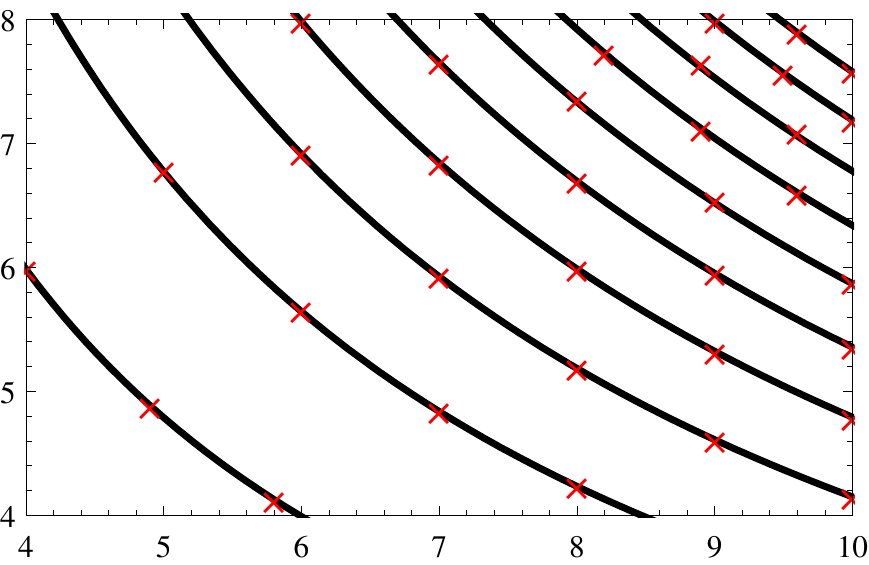}
			\put(-125,-10){$(2\pi\,B/\sqrt{\lambda})$}
			\put(2,145){\scriptsize{$\lambda=20$}}
			\put(-185,45){\colorbox{white}{\scriptsize{$7$}}}
			\put(-155,70){\colorbox{white}{\scriptsize{$14$}}}
			\put(-125,80){\colorbox{white}{\scriptsize{$22$}}}
			\put(-90,75){\colorbox{white}{\scriptsize{$29$}}}
			\put(-60,75){\colorbox{white}{\scriptsize{$36$}}}						\put(-60,93){\colorbox{white}{\scriptsize{$43$}}}
			\put(-36,89){\colorbox{white}{\scriptsize{$50$}}}
			\put(-38,105){\colorbox{white}{\scriptsize{$58$}}}
			\put(-18,115){\colorbox{white}{\scriptsize{$72$}}}
			\put(-193,15){\colorbox{white}{\scriptsize{$T=5$}}}
			\put(-220,70){\rotatebox[]{90}{$2\mu_5$}}\hspace{0.7cm}
			&
			\includegraphics[scale=0.8]{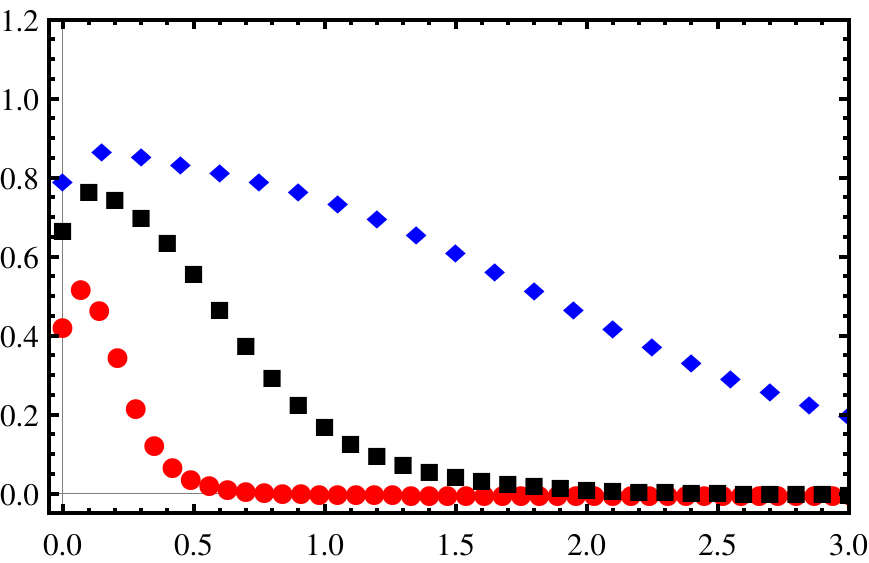}
			\put(-100,-10){$T$}
			\put(-150,20){\scriptsize{$b=1$}}
			\put(-130,38){\scriptsize{$b=10$}}
			\put(-60,60){\scriptsize{$b=100$}}				
			\put(-220,70){\rotatebox[]{90}{$\Delta_{10}(b,10,T)$}}
			\vspace{0.5cm}
			\\
			(a) & (b)
		\end{tabular}
		\caption{(a) The black curves are contours for $\Im I/\N$ at $T=5$ and red crosses indicates $B\mu_5=cte$ levels. (b) The temperature dependence of $\Delta_r(b,w,T)$ for three different values of magnetic field.  }
		\label{DeltaSym01}
	\end{center}
\end{figure}
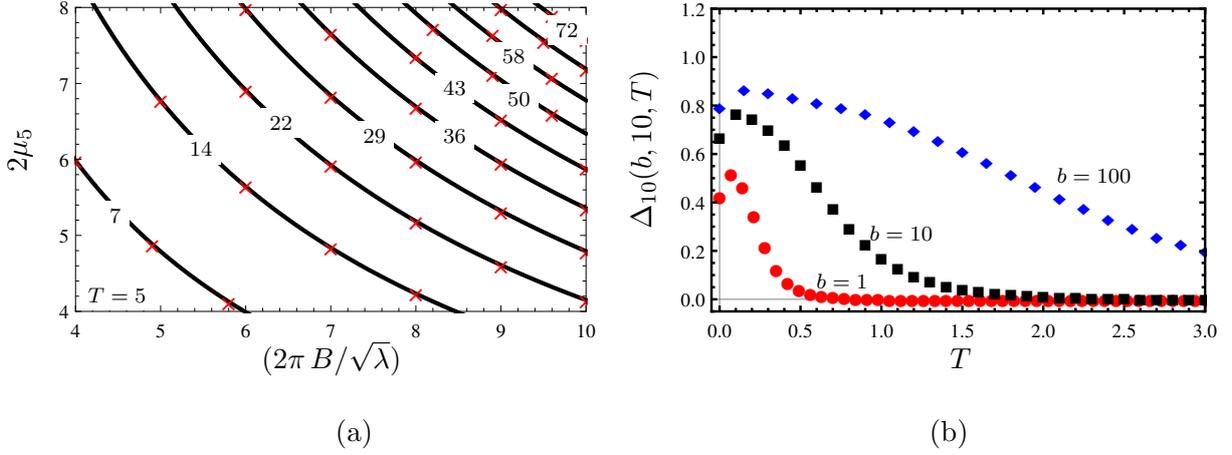

One can calculate $\Im I$ numerically for arbitrary magnetic field, axial chemical potential and temperature. The result is presented in a contour plot in Fig.\,\ref{DeltaSym01}(a) for $T=5$. The thick black curves are contours with constant values of $\Im I$ and red crosses on each curve shows the points with $B\,\mu_5=cte$. One sees that there is a good agreement between contour trades and $B\,\mu_5=cte$ which is a sign that $\Im I$ is a function of combination $B\,\mu_5$ at $T=5$ approximately.

Recalling the analytical results in different asymptotic values, it can be seen that in both cases of zero and non-zero temperature, the $\Im I$ is a function of $B\,\mu_5$ at $B\to 0$. It is not surprising because in \eqref{SimpAction} only combination $b\,w$ appears in the action  at $b\to 0$. Let us define the quantity
\bea
\Delta_r(b,w,T)=\frac{\Im I(r\,b,w/r,T)-\Im I(b,w,T)}{\Im I(r\,b,w/r,T)+\Im I(b,w,T)}.
\eea
If $\Im I$ is just a function of $b\,w$ (or $B\,\mu_5$) then the above quantity is equal to zero for all values of $r$. However, the non-zero $\Delta_r(b,w,T)$ for a given $r$ indicates that $\Im I$ is a function of $b$ and $w$ separately. As an example, $\Delta_{10}(0.1,10,5)\sim 10^{-6}$ while $\Delta_{10}(10,10,5)\sim 10^{-2}$. It means for small values of magnetic field this quantity approaches to zero which is compatible with previous analytical results for small values of $b$ in \eqref{LowBzeroT} and \eqref{LowBNonZeroT}.

Moreover, the dependence of $\Delta_r(b,w,T)$ with respect to the temperature leads to an interesting conclusion. This dependence is shown in Fig.\,\ref{DeltaSym01}(b) for $r=10$, $w=10$ and $b=1,\,10$ and $100$. According to this figure, for a given $B$ and $r$, the function $\Delta_r(b,w,T)$ approaches to zero for a large enough temperature. In other words, for finite values of $B$ and $\mu_5$ and large enough temperature $\Im I$ is approximately a function of combination $B\,\mu_5$.

\end{document}